\documentclass[10pt,letterpaper]{IEEEtran}
\usepackage{amsmath,amssymb,amsfonts,amsthm}
\usepackage{gensymb}
\usepackage{balance}
\usepackage{graphicx}
\usepackage{cite}
\usepackage[dvipsnames,usenames]{color}
\usepackage[colorlinks, linkcolor=Blue, filecolor =Red,citecolor=RoyalPurple]{hyperref}
\graphicspath{{figures/}}
\usepackage{scalerel}
\usepackage{nopageno}
\IEEEoverridecommandlockouts
\usepackage[lmargin=0.755in,rmargin=0.755in,tmargin=0.755in,bmargin=0.755in]{geometry}

\usepackage{adjustbox}
\usepackage[ruled,vlined]{algorithm2e}
\usepackage{algpseudocode}

\makeatletter
\def\endthebibliography{%
	\def\@noitemerr{\@latex@warning{Empty `thebibliography' environment}}%
	\endlist
}
\makeatother

 \usepackage{hyperref}
 \hypersetup{
     colorlinks=true,
     linkcolor=blue,
     filecolor=blue,
     citecolor = red,      
     urlcolor=cyan,
     }

\usepackage{hyperref}

\newcommand{\markedManu}{MARKED} 
\newcommand{\version}{arxiv} 

\newboolean{showArxiv}
\newboolean{showArxivAlt}

\ifthenelse{\equal{\version}{journal}}
{
	\setboolean{showArxiv}{false} 
	\setboolean{showArxivAlt}{true}
}

\ifthenelse{\equal{\version}{arxiv}}
{
	\setboolean{showArxiv}{true} 
	\setboolean{showArxivAlt}{false}
}

\ifthenelse{\equal{\markedManu}{MARKED}}
{

}

\ifthenelse{\equal{\markedManu}{nonMARKED}}
{

}

\def\plantFullname{District Cooling Energy Plant}
\def\plant{DCEP}
\def\plants{DCEPs}

\def\triangleq{\ensuremath{\stackrel{\Delta}{=}}}
\def\R{\mathbb{R}}

\def\qLref{\ensuremath{q^{L,\text{ref}}}}

\def\stateSet{{\sf{X}}}
\def\inputSet{{\sf{U}}}

\def\numPolImp{{\sf{N}}\ensuremath{_{\text{pol}}}}
\def\timeHorzRL{{\sf{T}}\ensuremath{_{\text{sim}}}}

\def\ct{\text{ct}}
\def\sw{\text{sw}}
\def\rw{\text{rw}}
\def\chw{\text{chw}}
\def\chws{\text{chw,s}}
\def\chwr{\text{chw,r}}
\def\bp{\text{bp}}
\def\ch{\text{ch}}
\def\cwr{\text{cw,r}}
\def\cw{\text{cw}}
\def\cws{\text{cw,s}}

\def\oawb{\text{oawb}}
\def\oa{\text{oa}}

\def\pw{\text{pw}}
\def\pump{\text{pump}}
\def\cond{\text{cond}}
\def\rej{\text{rej}}
\def\lw{\text{lw}}
\def\lws{\text{lw,s}}
\def\lwr{\text{lw,r}}
\def\indv{\text{indv}}

\def\qDotCTRej{\ensuremath{\dot{q}^{\ct, \text{rej}}}}
\def\qDotEvap{\ensuremath{\dot{q}^{\text{evap}}}}
\def\mDotCW{\ensuremath{\dot{m}^{\cw}}}

\def\TCWR{\ensuremath{T^{\cwr}}}
\def\TCWS{\ensuremath{T^{\cws}}}
\def\TOAWB{\ensuremath{T^{\text{oawb}}}}
\def\qChCapIndiv{\ensuremath{\dot{q}^{\ch,\indv}_{\mathrm{rated}}}}
\def\mDotLW{\ensuremath{\dot{m}^{\lw}}}
\def\mDotTW{\ensuremath{\dot{m}^{\tw}}}
\def\mDotSW{\ensuremath{\dot{m}^{\sw}}}
\def\mDotOA{\ensuremath{\dot{m}^{\oa}}}
\def\qDotLoad{\ensuremath{\dot{q}^L}}

\def\TChWS{\ensuremath{T^{\chws}}}
\def\TChWR{\ensuremath{T^{\chwr}}}
\def\TChWSMin{\ensuremath{T^{\chws}_{\text{min}}}}
\def\TCWR{\ensuremath{T^{\cwr}}}
\def\TCWRMax{\ensuremath{T^{\cwr}_{\text{max}}}}
\def\TCWS{\ensuremath{T^{\cws}}}

\def\twc{\text{twc}}
\def\tww{\text{tww}}
\def\tw{\text{tw}}
\def\nCh{\ensuremath{n^{\text{ch}}}}

\def\rd#1{{\color{red}{#1}}}

\def\pb#1{\notes{pb: \rd{#1}      }}

\newlength{\noteWidth}
\long\def\notes#1{\ifinner
	{\footnotesize #1}
	\else
	\marginpar{\parbox[t]{\noteWidth}{\raggedright\footnotesize #1}}
	\fi\typeout{#1}}

\usepackage{xparse}
\ExplSyntaxOn

\ExplSyntaxOff
\ExplSyntaxOn
\cs_new:Nn \pb_prop_gset_bykeys:Nn
{
	\prop_clear:N \l__pb_temp_prop
	\keys_set:nn { pb/propbykey } { #2 }
	\prop_gset_eq:NN #1 \l__pb_temp_prop
}
\cs_generate_variant:Nn \pb_prop_gset_bykeys:Nn { c }

\keys_define:nn { pb/propbykey }
{
	unknown .code:n = \prop_put:Nxn \l__pb_temp_prop { \l_keys_key_tl } { #1 }
}

\cs_generate_variant:Nn \prop_put:Nnn { Nx }

\NewDocumentCommand{\setupcollaborator}{mm}
{
	\prop_new:c { g_collaborator_#1_prop }
	\pb_prop_gset_bykeys:cn { g_collaborator_#1_prop } { #2 }
}

\NewDocumentCommand{\selectcollaborator}{m}
{
	\prop_map_inline:cn { g_collaborator_#1_prop }
	{
		\tl_set:cn { ##1 } { ##2 }
	}
}
\ExplSyntaxOff

\setupcollaborator{PBmac}
{
	NAME=Prabir Mac , laptop or mac mini,
	DiCEbibPATH=/Users/pbarooah/Dropbox/Dropbox/DiCElab_bib,
	ACbibPATH=/Users/pbarooah/Dropbox/Dropbox/austin_bib,
	NSRbibPATH=/Users/pbarooah/Dropbox/Dropbox/NSRbib,
	JBbibPATH =/Users/pbarooah/Dropbox/Dropbox/bibs,
}
\setupcollaborator{GZlinux}
{
	NAME=GZ on his lab computer,
	DiCEbibPATH=../../../DiCElab_bib,
}
\setupcollaborator{Austin}
{
	NAME=Austin on linux,
  	DiCEbibPATH=../../bibFiles/dicelab_bib,
	ACbibPATH=../../bibFiles/austin_bib,
	JBbibPATH=../../bibFiles/bibs,
}

\selectcollaborator{GZlinux}
\begin{document}
	\title{\vspace{0.25in} \centering Reinforcement Learning for Optimal Control of a \plantFullname}
        \author{
	\IEEEauthorblockN{Zhong Guo\IEEEauthorrefmark{1}$^,$\IEEEauthorrefmark{2}, Austin R. Coffman\IEEEauthorrefmark{1}, and Prabir Barooah\IEEEauthorrefmark{1}, \\\IEEEauthorrefmark{1}University of Florida.} 
	\thanks{\IEEEauthorrefmark{2} corresponding author, email: zhong.guo@ufl.edu.}
	\thanks{The authors are with the Dept. of Mechanical and Aerospace Engineering, University of Florida, Gainesville, FL 32601, USA. The research reported here has been partially supported by the NSF through award 1934322 (CMMI) and 2122313 (ECCS).}
	\vspace{-0.75cm}
}
\maketitle
\begin{abstract}
  District cooling energy plants (\plants) consisting of chillers, cooling towers, and thermal energy storage (TES) systems consume a considerable amount of electricity. Optimizing the scheduling of the TES and chillers to take advantage of time-varying electricity price is a challenging optimal control problem. The classical method, model predictive control (MPC), requires solving a high dimensional mixed-integer nonlinear program (MINLP) because of the on/off actuation of the chillers and charging/discharging of TES, which are computationally challenging. RL is an attractive alternative to MPC: the real time control computation is a low-dimensional optimization problem that can be easily solved. However, the performance of an RL controller depends on many design choices.

In this paper, we propose a Q-learning based reinforcement learning (RL) controller for this problem. Numerical simulation results show that the proposed RL controller is able to reduce energy cost over a rule-based baseline controller by approximately 8\%, comparable to savings reported in the literature with MPC for similar \plants. 
We describe the design choices in the RL controller, including basis functions, reward function shaping, and learning algorithm parameters. 
Compared to existing work on RL for \plants, the proposed controller is designed for continuous state and actions spaces.
	\end{abstract}
	
	\ifx 0
	\printnomenclature
	\nomenclature[CH]{$T_{CHWS}$}{chilled water temperature supplied from chillers($^\circ C$) }
	\nomenclature[CH]{$T_{CHWR}$}{chilled water temperature returned to chillers ($^\circ C$) }
	\nomenclature[CH]{$\dot{m}_{CHW}$}{chilled water flow rate (kg/sec) }
	\nomenclature[CH]{$P_{CH}$}{chiller power consumption (kW) }
	\nomenclature[CH]{$P_{CHWP}$}{power consumption of chilled water pumps (kW) }
	\nomenclature[CH]{$P_{tot}$}{total power consumption of chilled water plant (kW) }
	\nomenclature[CH]{$q_{L}$}{cooling required from loads (kW) }
	\nomenclature[CH]{$q_{CH}$}{cooling power of chiller (kW) }

	\nomenclature[CT]{$T_{CWS}$}{supply cooling water temperature ($^\circ C$) }
	\nomenclature[CT]{$T_{CWR}$}{return cooling water temperature ($^\circ C$) }
	\nomenclature[CT]{$T_{CW,makeup}$}{makeup water temperature ($^\circ C$) }
	\nomenclature[CT]{$\dot{m}_{CW}$}{cooling water flow rate (kg/sec) }
	\nomenclature[CT]{$\dot{m}_{CW,makeup/loss}$}{flow rate of makeup/lost cooling water (kg/sec) }
	\nomenclature[CT]{$P_{CT}$}{cooling tower power consumption (kW)}
	\nomenclature[CT]{$P_{CWP}$}{power consumption of cooling water pumps (kW) }
	\nomenclature[CT]{$T_{oawb}$}{ambient wet-bulb temperature ($^\circ C$) }
	\nomenclature[CT]{$\dot{m}_{oa}$}{ambient airflow rate (kg/sec) }
	\nomenclature[CH]{$q_{CT}$}{cooling power of cooling tower (kW) }
	
	\nomenclature[PA]{$C_{pw}$}{specific heat of water \big(kJ/(kg $\cdot^\circ C$)\big) }
	\nomenclature[PA]{$\rho_{w}$}{density of water (kg/m$^3$)}
	\nomenclature[PA]{$t_s$}{sampling period (min)}
	\nomenclature[PA]{$D^{\text{elec}}$}{Electricity cost ($\$/$(kWh))}
	\nomenclature[PA]{$\tilde{E}$}{aggregate thermal energy deviation (kWh)}
	\fi
	\section{Introduction}
\begin{figure}
	\centering
	\includegraphics[width=1\columnwidth]{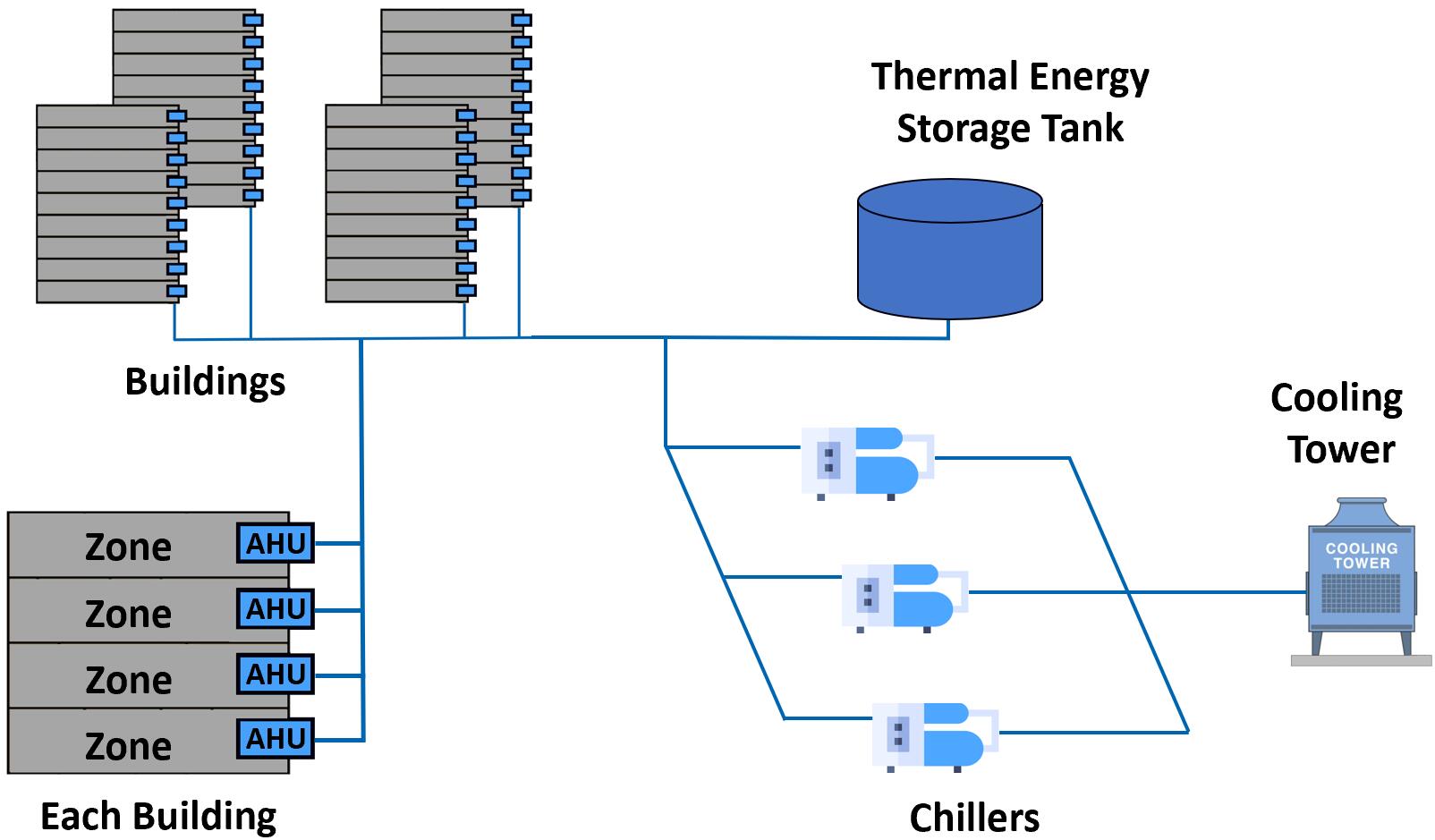}
	\caption{Layout of \plantFullname.}
	\label{fig:ChillerPlant_1}
\end{figure}

        In the U.S., 75\% of the electricity is consumed by buildings, and a large part of that is due to heating, ventilation, and air conditioning (HVAC) systems~\cite{CBECS:12}. In cities and campuses, a large part of the HVAC's share of electricity is consumed in \plantFullname s (\plants), also called central plants. A \plant\ produces and supplies chilled water to a group of buildings it serves (hence the moniker ``district''), and the air handling units in those buildings use the chilled water to cool and dehumidify air before supplying to building interiors.  Figure~\ref{fig:ChillerPlant_1} shows a schematic of such a plant, which consists of multiple chillers that produce chilled water, a cooling tower that rejects the heat extracted from chillers to the environment, and a thermal energy storage system (TES) for storing chilled water. Chillers - the most electricity intensive equipment in the \plant\ -  produces more chilled water than buildings' need when the electricity price is low and stored the extra chilled water in the TES. Then the chilled water in the TES can be used during periods of high electricity price to reduce the total electricity cost.

		At present,  \plants\ are typically operated with rule based control algorithms that use heuristics to reduce cost while meeting the load. But making the best use of the chillers and the TES to keep electricity cost at the minimum requires non-trivial decision making. Because of time coupling, especially due to the TES, the problem is best cast as an optimal control problem, of which the objective is the total electricity cost while meeting the thermal load from the buildings and equipment limitations are the constraints.

        A growing body of work has proposed algorithms for optimal real-time control of \plants. This includes Model Predictive Control (MPC), such as~\cite{RisbeckMixedintegerEnB:2017,rawlings2018economic,ColeMPCACC:2012,TouretzkyMPCJPS:2014, DengMPCASE:2014}. A twin challenge here is the discrete nature of certain decision variables, such as chiller on/off and TES charge/discharge commands, and the highly nonlinear dynamics of the equipment in \plants. This is a computationally challenging problem, which is typically addressed by avoiding it, such as by a MILP approximation. Solving large MILPs is also challenging, though.

        An alternative to MPC is Reinforcement Learning (RL): an umbrella term for a set of tools used to approximate an optimal policy from data collected from a physical system, or more frequently, its simulation. Despite a computationally burdensome learning phase, real-time control is simpler since control computation is an evaluation of a state-feedback policy. This advantage is particularly strong for problems involving discrete decision variables. RL is thus an attractive candidate for optimal control of \plants.

        In this paper we propose an RL  controller for a \plant\ with multiple chillers, a cooling tower and a TES, with a goal of reducing energy cost (over a rule-based control logic) while meeting the cooling load from buildings. The proposed controller uses a batch RL algorithm similar to the ``convex Q-learning'' proposed in recent work~\cite{ConvexLuACC:2021} and the classical least squares policy iteration (LSPI) algorithm~\cite{LeastSquaresLagoudakisJMLR:2003}. Closed loop simulations with a model calibrated using data from a campus in Singapore shows a cost saving around 8\%. This value is comparable to that of MPC controllers with mixed-integer formulation reported in the literature~\cite{RisbeckMixedintegerEnB:2017, DengMPCASE:2014}. Compared to MPC, the real time computation burden of the RL controller is trivial.

The baseline controller, that the proposed RL controller is compared to, has been designed to utilize the TES and time varying electricity prices (to the extent possible with heuristics) to reduce energy costs. The RL controller and baseline controller have the same information about the price to learn the trends in its time-variation: the current price and a backward moving average. 

Design of RL controllers for practically relevant applications with non-trivial dynamics is quite challenging. Firstly, a disadvantage of RL is that its performance depends on myriad design choices, not only on the stage cost/reward, function approximation architecture and bases, learning algorithm and method of exploration, but also on the choice of the state space itself. Secondly, training a RL controller is computationally intensive. Thirdly, and perhaps the most important for applications, if a particular set of design choices lead to a policy that does not perform well, there is no principled method to look for improvement. Although RL is being extensively studied in the control community, most works demonstrate their algorithms on plants with simple dynamics with a small number of states and inputs~\cite{DataDrivenBanjacCDC:2019,PolicyLuoToC:2017}. The model for a \plant\ used in this paper, arguably still simple compared to available simulation models (e.g.~\cite{FanOpensourceAE:2021}), is quite complex: it has 8-states, 5 control inputs, 3 disturbance inputs, and requires solving an optimization problem to compute the next state given the current state, control and disturbance.
A number of papers have investigated the use of RL for control of HVAC systems; see~\cite{ZhangBuildingBuildSys:2019,RamanReinforcement_ACC:2020,MasonReviewCCE:2019} and references therein. In particular, \cite{ZhangBuildingBuildSys:2019,RamanReinforcement_ACC:2020} propose RL controllers for air handling units in buildings. Dynamics of AHUs are considerably simpler than that of \plants. The refs. \cite{QiuModelfreeSTBE:2020,QiuModelfreeEB:2020,LiuEvaluationJSEE:2007} are more relevant to  the topic of this paper: control of a \plant\ with multiple chillers, cooling towers, and a TES system. In~\cite{QiuModelfreeSTBE:2020,QiuModelfreeEB:2020} RL controllers to operate chillers are proposed, and Watkin's Q-learning is used to train the controller with a discrete state space. The problem formulation is narrower; there is no TES, and the controller in \cite{QiuModelfreeSTBE:2020} computes only one setpoint, the chilled water supply water temperature setpoint, while~\cite{QiuModelfreeEB:2020} computes two setpoints, operating frequencies of cooling tower fans and cooling water pumps. Another relevant work \cite{LiuEvaluationJSEE:2007} also uses Q learning to train an controller that determines zone temperature setpoints and TES charging/discharging flow rates. However, trajectories of external inputs, e.g., outside air temperature and electricity price, are the same for all training days in~\cite{LiuEvaluationJSEE:2007}. Therefore, the RL controller needs to be retrained every time when distinct external input trajectories are used. 

The contributions of this work over the related literature on ``RL for \plants'' cited above are as follows. One, the proposed RL controller does not require discretizing the state space like the prior works~\cite{QiuModelfreeSTBE:2020,QiuModelfreeEB:2020}. Two, it computes all five setpoints required to operate a \plant\ (described in detail in Section~\ref{sec:sysDesc}), compared to one or two as done in the prior works. Three, comparing to~\cite{LiuEvaluationJSEE:2007}, we treat external inputs as time-varying disturbances and include them as RL states, making the proposed RL controller valid for any time-varying disturbances, not just the trajectories considered in training. 

The rest of the manuscript is organized as follows. Section~\ref{sec:sysDesc} describes the \plantFullname\ and its simulation model as well as the control problem. Section~\ref{sec:RL} describes the proposed controller, and Section~\ref{sec:eval} provides simulation evaluation of the controller. Section~\ref{sec:conclusion} concludes the paper.
	\section{System description and control problem}\label{sec:sysDesc}
\begin{figure*}[h]
		\centering
		\includegraphics[width=1.75\columnwidth, height=0.82\columnwidth]{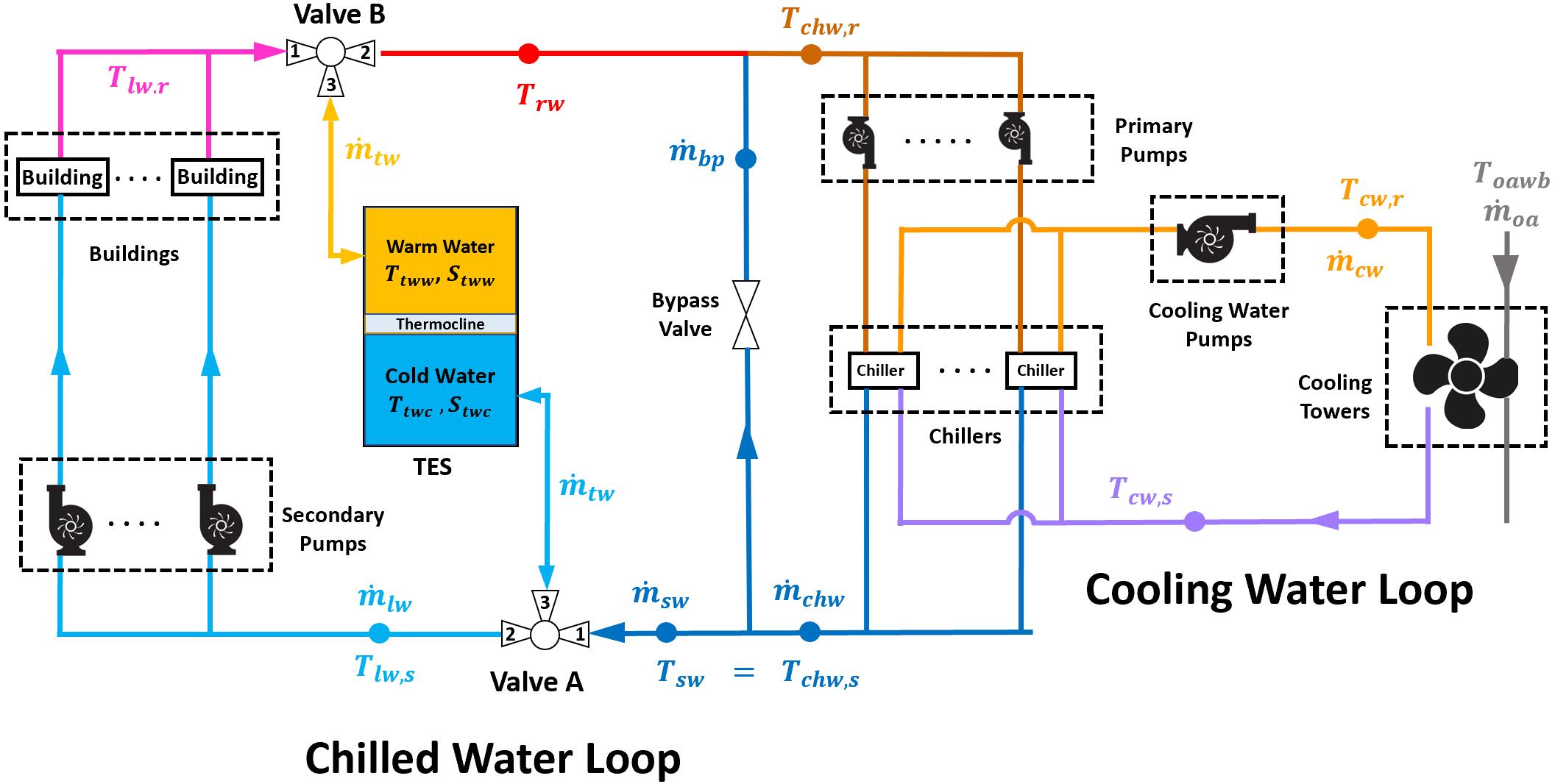}
		\caption{Detailed description of \plantFullname.}
		\label{fig:ChillerPlant}
	\end{figure*}

	The \plant\ contains multiple chillers and chilled water pumps, a cooling tower and cooling water pumps, and finally a collection of buildings that uses the chilled water to provide air conditioning; see Figure~\ref{fig:ChillerPlant}. 
	The heat load from the buildings is absorbed by the cold chilled water supplied by the \plant, and thus the return chilled water temperature is warmer. This part of the water is called \emph{load water}, and the related variables are denoted by superscript $\lw$ for ``load water''; see Figure~\ref{fig:ChillerPlant}. The chiller loop (subscript $\ch$) removes this heat and transmits it to the cooling water loop (subscript $\cw$). The cooling water loop absorbs this heat  and sends it to the cooling tower, where this heat is then rejected to the ambient. The cooling tower will cool down the cooling water returned from the chiller by passing both the sprayed cooling water and ambient air through a fill. During this process, a small amount of water spray will evaporate into the air, removing heat from the cooling water. The cooling water loss due to its evaporation is replenished by fresh water, thus we assume the supply water flow rate equals to the return water flow rate at the cooling tower. A fan or a set of fans is used to maintain the ambient airflow at the cooling tower.

        Connected to the chilled water loop is a TES tank that stores water (subscript $\tw)$. The total volume of the water in the TES tank is constant, but a thermocline separates two volumes: cold water that is supplied by the chiller (subscript $\twc$ for ``tank water, cold'') and warm water returned from the load (subscript $\tww$ for ``tank water, warm'').
\subsection{\plant\ dynamics}\label{sec:plantModel}
	Time is discretized with sampling period $t_s$. A counter $k=0,1,\dots,$ denotes the time step. The control command for the \plant\ at time $k$ is:
	\begin{align} \label{eq:inputVecPlant}
	u_k = [n^{\ch}_k, \ \dot{m}^{\lw}_k, \ \dot{m}_k^{\tw}, \ \dot{m}^{\cw}_k, \dot{m}_k^{\oa}]^T \in \inputSet,
	\end{align}
        where $n^{\ch}$ is the number of chillers on, and $\dot{m}$ is mass flow rate of water, with superscripts describing the name of a water loop, except  $\dot{m}^{\oa}$, which is the mass flow rate of air passing through the fill at the cooling tower. Each of these variables can be independently chosen as setpoints since lower level PI-control loops maintain them. There are limits to these setpoints, which determine the admissible input set $\inputSet$:
	\begin{align} \nonumber
	\inputSet &\triangleq  \{0,\dots, n_\text{max}^{\ch}\} \times [\dot{m}_\text{min}^{\lw},\dot{m}_\text{max}^{\lw}] \times [\dot{m}_\text{min}^{\tw},\dot{m}_\text{max}^{\tw}] \dots \\ \times &[\dot{m}_\text{min}^{\cw},\dot{m}_\text{max}^{\cw}] \times [\dot{m}_\text{min}^{\oa},\dot{m}_\text{max}^{\oa}]\subset \R^5.
	\end{align}
The state of the \plant\ is $x^p$ (superscript $p$ is for plant):
        \begin{align}
\label{eq:stateVecPlant}
x^p_k \triangleq[T_k^{\lwr}, s_k^\tww, s_k^\twc, T_k^\twc, T_k^\tww,T_k^{\chws}, T_k^{\cwr}, T_k^{\cws}]^T,
        \end{align}
        where $s^\tww, s^\twc$ are the volumes of the warm water and cold water in the TES tank. The other state variables are temperatures at various locations - both supply (subscript ``$,s$'') and return (subscript ``$,r$") - in the water loops: load water, cooling water, tank water, and chiller; See Figure~\ref{fig:ChillerPlant}. The plant state $x^p$ is affected by exogenous disturbances $w^p_k:=[T^{\oawb}_k,q^{L,\text{ref}}_k]^T \in \R^2$, where $q^{L,\text{ref}}_k$ is the thermal load (heat from the buildings the load water needs to remove), and $T^{\oawb}_k$ is the ambient wet bulb temperatures. 
        
        The control command and disturbances affect the state through a highly nonlinear dynamic model:
      \begin{align}\label{eq:plant-model-f}
	x^p_{k+1} & = f(x^p_k,u_k,w^p_k),
      \end{align}
  	\ifshowArxivAlt
        that is described in the Appendix in~\cite{GuoReinforcementReport:2021}. 
	\fi
	\ifshowArxiv
		that is described in the Appendix. 
	\fi
	The dynamics~\eqref{eq:plant-model-f} are implicit: current output values depend upon next state values, since each heat exchanger only has a limited capacity.  Hence it requires some form of iterative updates to simulate the dynamics, e.g.,  the method developed in~\cite{YuOptimizationAE:2008}. A generalized way to perform the iterative update to account for the limits of heat exchange capacities is by solving a constrained optimization problem, which is the method used in this work. 
\subsection{Electrical demand and energy cost}\label{sec:Pelec-formulas}
	In the \plant\ considered, the only energy used is electricity. The relationship between the thermal quantities and the electrical demand in chillers and cooling tower are complex. We model the chillers power consumption $P^{\ch}$ as~\cite{ASHRAE:guideline14_2002}: 
	\begin{align}\label{eq:P_CH}
          P^{\ch}_k & = (\frac{T^{\cws}_k}{T^{\chws}_k}-1)q^{\ch}_k-\beta_1+\beta_2T_k^{\cws}-\beta_3\frac{T_k^{\cws}}{T_k^{\chws}}.
	\end{align}
	Power consumption of water pumps is modeled using the black-box model in~\cite{RisbeckMixedintegerEnB:2017}:
	\begin{align}
	P^{\chw,\pump}_k &= \alpha_1\ln(1+\alpha_2\dot{m}_k^{\chw})+\alpha_3\dot{m}_k^{\chw}+\alpha_4, \label{eq:P_chwPumps} 
	\\P_k^{\cw,\pump} &= \gamma_1 \ln(1+\gamma_2\dot{m}_k^{\cw})+\gamma_3\dot{m}^{\cw}_k+\gamma_4.\label{eq:P_cwPumps} 
	\end{align}
	Finally, the electrical power consumption of the cooling tower mainly comes from its fan and is modeled as~\cite{Braun:ASHRAE_Trans_1996}:
	\begin{align} \label{eq:P_CT_mDot_oa}
	P^{\ct}_k = \lambda(\dot{m}^{\oa}_k)^3.
	\end{align}
	The constants $\alpha_i,\beta_i,\gamma_i$, and $\lambda$ are empirical parameters.

        \subsection{Model calibration and validation} \label{sec:modelCal}
\begin{figure}[h]
		\centering
		\includegraphics[width=1\linewidth]{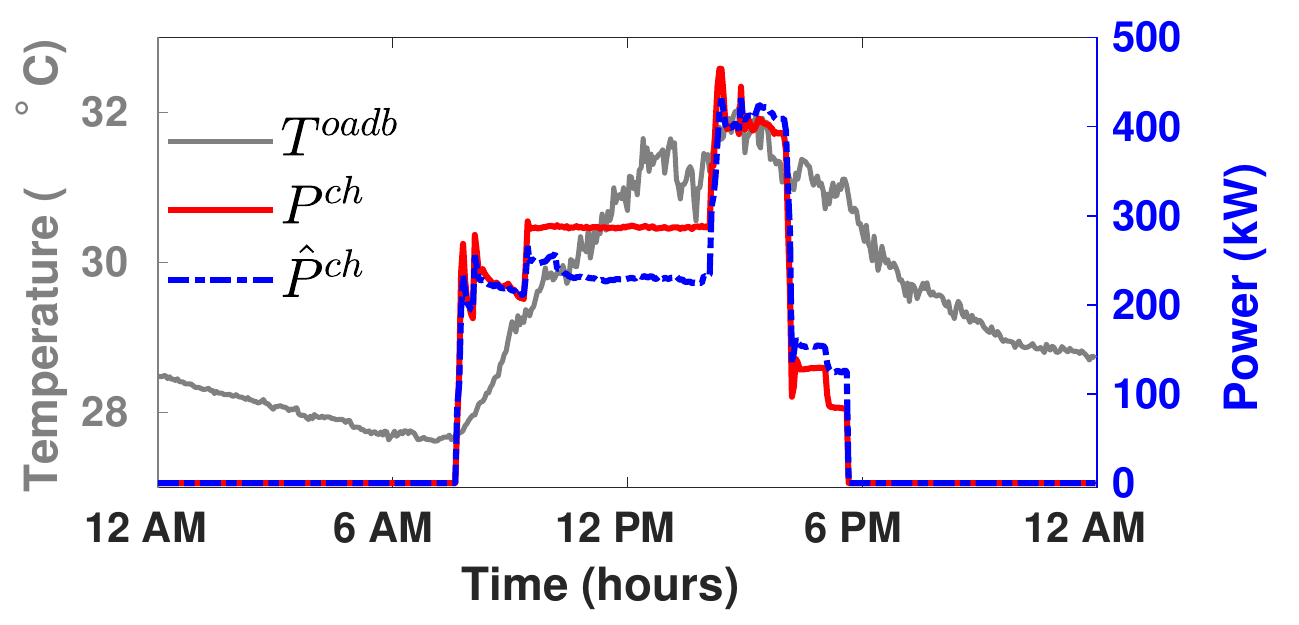}
		\caption{Out of sample prediction for $P^{\ch}$ using the identified weights for the model~\eqref{eq:P_CH}.}\label{fig:Iden_P_CH}
	\end{figure}

        The parameters of the simulation model in Section~\ref{sec:plantModel} and electrical demand model in Section~\ref{sec:Pelec-formulas} are described in  	
        \ifshowArxivAlt
        the Appendix in~\cite{GuoReinforcementReport:2021}. 
        \fi
        \ifshowArxiv
        the Appendix. 
        \fi 
        They are identified using data from the energy management system in United World College of South East Asia Tampines Campus in Singapore~\cite{MillerSingporeDatabaseUrl:2014,MillerSingporeDatabasePaper:2014}. We use 80\% of data for model identification and 20\% of data for verification. The out-of-sample prediction results for the total electrical demand are shown in Figure~\ref{fig:Iden_P_CH}. Comparison between data and prediction for other variables are not shown due to lack of space.
    \subsection{The control problem} \label{sec:contGoal}
	The control goal is to operate the \plant\ in a way so that the cost is minimized while the cooling load $\qLref_k$ is met. The total electric power consumption of the \plant\ is:
	\begin{align}
		P^{\text{tot}}_k= P^\ch_k + P^\ct_k + P^{\chw, \pump}_k + P^{\cw, \pump}_k, \label{eq:P_tot}
	\end{align}
	so that the electricity cost at time $k$ is 
	\begin{align} \label{eq:elecCost}
		c^\text{E}_k =  t_s\rho_k P^{\text{tot}}_k,
	\end{align}
	where $\rho_k$ $(\frac{\text{USD}}{\text{kWh}})$ is the electricity price. The control problem for a scheduling horizon $T$ is then:
	\begin{align} \label{eq:OC_plant}
		\min_{\{u_k\}_{k=0}^{T-1}} \ &\sum_{k=0}^{T-1}c_k^{\text{E}} \\
		\text{s.t.} \quad &x^p_{k+1} = f(x^p_k,u_k,w^p_k), \ x^p_0 = x, \nonumber\\
		&q^L_k(x^p_k,u_k) = q^{L,\text{ref}}_k \ \text{and} \ u_k \in \inputSet(x^p_k,w_k).	\nonumber
	\end{align}	
where $q^L_k(x^p_k,u_k) $ is the actual cooling load met by the \plant, which is a complicated function of the states and inputs, and 
	\begin{align}  \label{eq:stateInputSet}
		\begin{split}
		\inputSet(x^p_k,w_k) \triangleq \big\{ &u_k \in \inputSet: \
		\mDotLW_k \geq \mDotLW_{\text{min}}, \ \qDotCTRej_k \geq \qDotEvap_k,\\
		&\mDotLW_k+\mDotTW_k \leq \dot{m}^{\chw}_k, \\ 
		&T^{\cws,-}_{k+1} \leq T^{\cws}_{k+1} \leq T^{\cws,+}_{k+1}, \\
		&S^{\tw}_{\min}\leq S^{\twc}_{k} + t_s\mDotTW_k \leq S^{\tw}_{\max}
		\big\},
			\end{split}	
	\end{align}
	where  $T^{\cws,-}_{k+1} = \max\big(25,T^{\oawb}_{k+1}+0.5\big)$ and  $T^{\cws,+}_{k+1} = \min\big(T^{\cwr}_k , 40\big)$.

    The optimization problem~\eqref{eq:OC_plant} is an MINLP due to $n^{\ch}_k$ being an integer and the nonlinear dynamics~\eqref{eq:plant-model-f}. The solution of~\eqref{eq:OC_plant} provides the best achievable control performance over the horizon $T$: only an oracle equipped with a perfect prediction of disturbance over the whole planning horizon will be able to compute it. With imperfect forecasts, one could use MPC to bring robustness to forecast uncertainty.  Still, the challenge of computing the MINLP is a severe one. The ref.~\cite{RisbeckMixedintegerEnB:2017} used piece-wise linearization leading to an MILP. In the following we propose a RL controller that does not need to solve even MILPs. 

        All the plant state variables $x^p$ can be measured with sensors. The same for the disturbances, except there is no sensor for the cooling load $\qLref_k$. There is a large literature on estimating and/or forecasting loads for buildings; see~\cite{bracha:2002,GuoAggregationEnB:2020} and references therein. We therefore assume $\qLref_k$ is known to the controller at $k$.
	\section{RL basics and proposed controller}\label{sec:RL}
	
	\subsection{RL basics} \label{sec:RLframe}
	For the following construction, let $x$ represent the state with state space $\stateSet$ and $u$ the input with input space $\inputSet(x)$. Now consider the following infinite horizon discounted optimal control problem
	\begin{align}
	J^*(\bar{x}) = &\min_{\mathbf{U}} \quad \sum_{k=0}^{\infty}\gamma^k c(x_k,u_k), \quad x_0 = \bar{x}, \label{eq:OC_cost} \\
	&\text{s.t.} \quad  x_{k+1} = F(x_k,u_k), \ u_k \in \inputSet(x_k), \nonumber
	\end{align}
	where $\mathbf{U}\triangleq\{u_0,\dots, \}$, $c: \stateSet \times \inputSet \to \R^{\geq 0}$ is the stage cost, $\gamma \in (0,1)$ is the discount factor, $F(\cdot,\cdot)$ defines the dynamics, and $J^* : \stateSet \rightarrow \mathbb{R}^+$ is the optimal value function. The goal of the RL framework is to learn an approximate optimal policy $\phi : \stateSet \rightarrow \inputSet$ for the problem~\eqref{eq:OC_cost} without requiring explicit knowledge of the model $F(\cdot,\cdot)$. The learning process is based on the $Q$ function. Given a policy $\phi$ for the problem~\eqref{eq:OC_cost}, the $Q$ function associated with this policy is defined as
	\begin{align} \label{eq:QfuncFixPol}
	Q_\phi(x,u) = \sum_{k=0}^{\infty}\gamma^k c(x_k,u_k), \quad x_0 = x, \quad u_0 = u,
	\end{align}
	where for $k\geq 0$ we have $x_{k+1} = F(x_k,u_k)$ and for $k \geq 1$ we have $u_k = \phi(x_k)$. 
	A well known fact is that the optimal policy satisfies~\cite{sutbar18}
	\begin{align} \label{eq:argMinQ}
	\phi^*(x) = \arg\min_{u\in\inputSet(x)} \ Q^*(x,u), \quad \text{for all} \quad x\in\stateSet,
	\end{align}
	where $Q^* \triangleq Q_{\phi^*}$ is the $Q$ function for the optimal policy. Further, for any policy $\phi$ the $Q$ function satisfies the following fixed point relation
	\begin{align} \label{eq:fixPolBellEq}
	Q_\phi(x,u) = c(x,u) + \gamma Q_\phi\big(x^+, \phi(x^+)\big),
	\end{align}
	for all $u\in\inputSet(x)$ and $x\in\stateSet$ and $x^+ = F(x,u)$. The above relation is termed here as the fixed-policy Bellman equation.

	\subsection{RL algorithm: data driven policy iteration} \label{sec:RLalg}
	The learning algorithm has two parts: policy evaluation and policy improvement. First, in policy evaluation, a parametric approximation to the fixed policy $Q$ function is learned by constructing a residual term from~\eqref{eq:fixPolBellEq} as an error to minimized. Second, in policy improvement, the learned approximation is used to define a new policy based on~\eqref{eq:argMinQ}. For policy evaluation, suppose for a policy $\phi$ the $Q$ function is approximated as
	\begin{align} \label{eq:qHatTheory}
	Q^\theta_\phi(x,u) \approx \hat{q}_\theta(x,u)
	\end{align}
	where $\hat{q}_\theta(\cdot,\cdot)$ is the function approximator (e.g., a neural network) and $\theta \in \mathbb{R}^d$ is the parameter vector (e.g., weights of the network). 	To fit the approximator, suppose that the system is simulated for $\timeHorzRL$ time so that $\timeHorzRL$ tuples of $(x_k,u_k,x_{k+1})$ are collected to produce $\timeHorzRL$ values of 
	\begin{align}
		d_k(\theta) = c(x_k,u_k) + \gamma \hat{q}_\theta(x_{k+1},\phi(x_{k+1})) - \hat{q}_\theta(x_k,u_k),
	\end{align} 
	which is the temporal difference error for the approximator.
	We then obtain $\theta^*$ by solving the following optimization problem
		\begin{align} \label{eq:polEval}
		\begin{split}
			\theta^* \triangleq \arg\min_\theta \  &\|D(\theta)\|_2 + \alpha \|\theta - \bar{\theta}\|_2,
		\end{split}
	\end{align}
	where $D(\theta) \triangleq [d_0(\theta), \dots, d_{\timeHorzRL-1}(\theta)]$. The term $\|\theta - \bar{\theta}\|_2$ is a regularizer and $\alpha$ is a gain. The values of $\bar{\theta}$ and $\alpha$ are specified in step 3) of Algorithm~\ref{alg:dataDrivePolIter}. The solution to~\eqref{eq:polEval} results in $Q^{\theta^*}_\phi$, which is an approximation to $Q_\phi$. The quantity $Q^{\theta^*}_\phi$  can be used to obtain an improved policy, denoted $\phi^+$, through
	\begin{align} \label{eq:polUpdate}
	\phi^+(x) = \arg\min_{u\in\inputSet(x)}\ Q^{\theta^*}_\phi(x,u), \quad \text{for all} \quad x \in \stateSet.
	\end{align}
	This process of policy evaluation~\eqref{eq:polEval} and policy improvement~\eqref{eq:polUpdate} can be repeated. This iterative procedure is described formally in Algorithm~\ref{alg:dataDrivePolIter}.  
	\begin{algorithm} 
		\SetAlgoLined
		\KwResult{An approximate optimal policy $\phi^{\numPolImp}(x)$. }
		\KwIn{$\timeHorzRL$, $\theta^0$, $\numPolImp$, $\beta > 1$}

		\For{$j=0,\dots, \numPolImp-1$ }{
			\medskip
			 
			\textbf{1)} Obain input sequence $\{u^j_k\}_{k=0}^{\timeHorzRL-1}$, initial state $x^j_0$, and state sequence $\{x^j_k\}_{k=1}^{\timeHorzRL}$. \ 
			
			\medskip
			
			\textbf{2)} For $k=1,\dots, \timeHorzRL$, obtain: $\phi^{j}(x_k) = \arg\min_{u\in\inputSet(x_k^j)} \hat{q}_{\theta^j} (x^j_k,u)$. \
			
			\medskip
			
			\textbf{3)} Set $\bar{\theta} = \theta^j$ and $\alpha = \frac{j}{\beta}$ appearing in~\eqref{eq:polEval}. 
			
			\textbf{4)} Use the samples $\{u^j_k\}_{k=0}^{\timeHorzRL-1}$, $\{x^j_k\}_{k=0}^{\timeHorzRL}$, and $\{\phi^{j}(x_k)\}_{k=1}^{\timeHorzRL}$ to construct and solve~\eqref{eq:polEval} for $\theta^*$.
			
			\medskip
			
			\textbf{5)} Set $\theta^{j+1} = \theta^*$.
			
			\medskip
	}
		\caption{Data Driven Policy Iteration: Batch mode and off-policy}
		\label{alg:dataDrivePolIter}
	\end{algorithm}	

	This algorithm is inspired by: (i) the Batch Convex-Q learning algorithm found in~\cite[Section III]{ConvexLuACC:2021} and (ii) the least squares policy evaluation (LSPI) algorithm~\cite{LeastSquaresLagoudakisJMLR:2003}. The approach here is simpler than the batch optimization problem that underlies the algorithm in~\cite[section III]{ConvexLuACC:2021}, which has an objective function that itself contains an optimization problem. In comparison to~\cite{LeastSquaresLagoudakisJMLR:2003} we include a regularization term and constraint to ensure the Q function approximate is non-negative.  
\subsection{Proposed RL controller for \plant} \label{sec:RLarch}
         	We now specify the ingredients required to apply Algorithm~\ref{alg:dataDrivePolIter} to obtain a RL controller (i.e., a state feedback policy) for the \plant\ from simulation data. Namely, (1) the state description, (2) the cost function design, (3) the approximation architecture, and (4) the exploration strategy. Parts (1), (2), and (3) refer to the setup of the optimal control problem the RL algorithm is attempting to approximately solve. Part (4) refers to the selection of how the state/input space is explored (step 1 in Algorithm~\ref{alg:dataDrivePolIter}).
	
	\subsubsection{State space description}
	In RL, the construction of the state space is an important feature, and the state is not necessarily the same as the plant state. To define the state space for RL, we first denote $w_k$ as the vector of exogenous variables
	\begin{align}
		w_k = [(w_k^p)^T, \rho_k,\bar{\rho}_k] \in \R^4.
	\end{align}
	where $\bar{\rho}_k = \frac{1}{\tau}\sum_{t = k - \tau}^{k}\rho_t$
	is a backwards moving average of the electricity price with $\tau$ chosen to represent $4$ hours.
     The expanded state for RL is:
	\begin{align}
		x_k \triangleq [x_k^p, w_k]^T \in \stateSet \triangleq \subset \R^{12}. 
	\end{align}
    Notice that the set~\eqref{eq:stateInputSet} can now be written as:
    \begin{align}
    		\inputSet(x_k) = \inputSet(x_k^p,w_k). 
    \end{align}
Note that \emph{ a state feedback policy is implementable} since all entries of $x_k$ can be measured with commercially available sensors (e.g., outside wet-bulb temperature, $T_{oawb}$), or estimated from measurements (e.g., the thermal load from buildings, $\qLref$), or known via real-time communication (e.g., the electricity prices, $\rho_k$ and $\bar{\rho}_k$).

\subsubsection{Design of stage cost}
    The design of the stage cost is an important aspect of RL. We wish to obtain a policy that tracks the load  $q^{L,\text{ref}}_k$ whilst spending minimal amount of money, as described in section~\ref{sec:contGoal}. So we choose
        \begin{align}
          \label{eq:5}
           c(x_k,u_k) \triangleq c_k^{\text{E}} + \kappa\left(q^L_k - q^{L,\text{ref}}_k\right)^2,
        \end{align}
    where $\kappa \gg 1$ to prefer load tracking over energy cost.
    
    \subsubsection{Approximation architecture}
  We choose the following linear-in-the-parameter approximation of the $Q$ function:
    \begin{align} \label{eq:QfuncApprox}
    Q^{\theta}_{\phi}(x,u) \approx \hat{q}_{\theta}(x,u) = \sum_{\ell=1}^d\psi_\ell(x,u)\theta_\ell,
    \end{align}
    where $\psi_\ell(x,u)$ are basis functions and $\theta \in \R^d$ is the parameter vector. We elect a quadratic basis, so that each $\psi_\ell(x,u)$ is of the form $xu$, $x^2$, or $u^2$. A subset of all possible combinations are included in the basis. We can equivalently express the approximation~\eqref{eq:QfuncApprox} as
    \begin{align} \label{eq:QfuncMatForm}
    	Q^{\theta}_{\phi}(x,u) = [x,u]P_\theta [x,u]^T,
    \end{align}
    for appropriately chosen $P_\theta$. In this form it is straightforward to ensure the Q function non-negative by enforcing the convex constraint $P_\theta \geq 0$.

	\subsubsection{Exploration strategy} \label{sec:simEnv}
	Exploration refers to how the state/input sequences appearing in step 1) of Algorithm~\ref{alg:dataDrivePolIter} are simulated. We utilize a modified $\epsilon-$greedy exploration scheme. At time step $k$ of iteration $j$, we obtain the input $u_k^j$ from one of three methods: (i) by using the policy in step 2) of Algorithm~\ref{alg:dataDrivePolIter}, (ii) electing uniformly random feasible inputs, and (iii) using a rule-based baseline controller (described in Section~\ref{sec:ruleBased}). The states are obtained sequentially through simulation, starting from state $x^j_0$ for each $j$. The choice to use either of the three controllers is determined by the probability mass function $\nu^j_{\text{exp}} \in \R^3$, which depends on the iteration index of the policy iteration loop: 	
	\begin{align}
		\nu^j_{\text{exp}} = \begin{cases}
			[0,0.1,0.9] & \text{for} \ j\leq 5. \\
			[0.5,0.25,0.25] & \text{for} \ j> 5.
		\end{cases}
	\end{align}
	The entries correspond to the probability of using the corresponding control strategy, which appear in the (i)-(iii) order as just introduced. The rational for this choice is that the BL controller provides ``reasonable'' state input examples for the RL algorithm in the early learning iterations so to steer the parameter values in the correct direction. After this early learning phase, weight is shifted towards the current working policy so to force the learning algorithm to update the parameter vector in response to its actions.
        \subsection{Real time implementation}
Once the RL controller is trained, it computes the control command in real-time as
	\begin{align} \label{eq:uk-argMinQ}
u_k := 	\phi^*(x_k) = \arg\min_{u\in\inputSet(x_k)} \ Q^{\hat{\theta}}(x_k,u),
	\end{align}
where $\hat{\theta}$ is the parameter vector learned at the end of the learning phase - after $\numPolImp$ policy improvement steps -  that is described above. The value of $d$  used in the numerical results reported later is $d = 35$.  Due to non-convexity of the set $\inputSet(x_k)$ and $\nCh_k$ being an integer variable, the problem~\eqref{eq:uk-argMinQ} is non-convex. To solve it, for each possible value of $\nCh_k$, we solve the corresponding continuous variable non-linear program using CasADi/IPOPT~\cite{Andersson2019,wacbie:2006}, and then choose the minimum out of $\nCh_{\text{max}}$ solutions.

	\section{Performance evaluation}\label{sec:eval}
	\subsection{Rule-based Baseline Controller}\label{sec:ruleBased} In order to evaluate the performance of the RL controller, we will compare it to a rule-based baseline controller (BL). The baseline controller determines the setpoints $u_k = [n^{\ch}_k, \ \dot{m}^{\lw}_k, \ \dot{m}_k^{\tw}, \ \dot{m}^{\cw}_k, \dot{m}_k^{\oa}]^T$ as follows.  The quantity \mDotLW\ is determined based on the nominal temperatures of chilled water in the cooling coil (7 and 12 Celsius for $\TChWS$ and $\TChWR$, respectively). The water flow rate for the TES, \mDotTW, is chosen based a comparison between the current electricity price and a four-hour backward moving average electricity price. If the current electricity price is cheaper than its four-hour backward average, the TES is charged at maximum flow rate. Similarly, if the current electricity price is higher than its four-hour backward average, the TES is discharged at maximum flow rate. TES will stop charging/discharging once either the cold water or the warm water in the TES reaches its maximum/minimum bounds. The BL controller then determines the number of chillers \nCh\ based on the required flow rates specified from the choices of \mDotLW\ and \mDotTW. The number of chillers \nCh\ and cooling load \qLref\ then together dictate the required value of the cooling water flow rate \mDotCW. The value of outside air flow rate affecting the cooling tower is determined based on the inlet conditions of the cooling tower so to ensure proper operation.
	\ifx
\begin{table}
	\centering
	\caption{Simulation Parameters}
	\setlength{\arrayrulewidth}{0.05cm}
	\begin{tabular}{ l c c | l c c} 
		\hline 
		Parameter & Unit & value & Parameter & Unit & value \\ \hline
		$\beta$ & N/A & 100 & $\gamma$ & N/A & 0.97 \\
		$\timeHorzRL$ & N/A & 432 & $\numPolImp$ & N/A & 50 \\
		$t_s$ & minutes & 10 & $d$ & N/A & 35 \\
		$\kappa$ & N/A & 500 & $\theta_0$ & N/A & random \\
		$\frac{\tau t_s}{60}$ & hours & 4 & $\tau$ & N/A & 24 \\
		$n^{\ch}_\text{max}$ & N/A & 7 & $\mDotTW_{\text{max}}/\mDotTW_{\text{min}}$ & $\frac{\text{kg}}{\text{sec}}$ & 30/-30 \\
		$\mDotLW_\text{max}/\mDotLW_\text{min}$ & $\frac{\text{kg}}{\text{sec}}$ & 350/20 & $\mDotCW_{\text{max}}/\mDotCW_{\text{min}}$ & $\frac{\text{kg}}{\text{sec}}$ & 300/20 \\
		$ \frac{S^{\twc/\tww}_{\text{max}}}{\rho_{w}} $ & $m^3$ & 900 & $ \frac{S^{\twc/\tww}_{\text{min}}}{\rho_{w}} $ & $m^3$ & 45  \\
		\hline
	\end{tabular}
*$\rho_{w}$ is the density of water (kg/$m^3$).
	\label{tab:simParams}
\end{table}
\fi

\begin{table}
		\centering
		\caption{Simulation Parameters}
		\setlength{\arrayrulewidth}{0.05cm}
		\begin{tabular}{ l c c | l c c}
				\hline
				Parameter & Unit & value & Parameter & Unit & value \\ \hline
				$\beta$ & N/A & 100 & $\gamma$ & N/A & 0.97 \\
				$\timeHorzRL$ & N/A & 432 & $\numPolImp$ & N/A & 50 \\
				$t_s$ & minutes & 10 & $d$ & N/A & 35 \\
				$\kappa$ & N/A & 500 & $\theta_0$ & N/A & random \\
				$\tau$ & hours & 4 &  $\frac{S^{\twc}_{\text{max}}}{\rho_{w}}/\frac{S^{\twc}_{\text{min}}}{\rho_{w}} $ & $m^3$ & 45/900 \\
				$n^{\ch}_\text{max}$ & N/A & 7 & $\mDotTW_{\text{max}}/\mDotTW_{\text{min}}$ & $\frac{\text{kg}}{\text{sec}}$ & 30/-30 \\
				$\mDotLW_\text{max}/\mDotLW_\text{min}$ & $\frac{\text{kg}}{\text{sec}}$ & 350/20 & $\mDotCW_{\text{max}}/\mDotCW_{\text{min}}$ & $\frac{\text{kg}}{\text{sec}}$ & 300/20\\
				\hline
		\end{tabular}
		*$\rho_{w}$ is the density of water (kg/$m^3$).
		\label{tab:simParams}
\end{table}
\subsection{Simulation setup}
Closed loop simulations are performed with the RL and BL controllers, with the plant being the dynamic model~\eqref{eq:plant-model-f}.
The weather data is a part of the Singapore data set described in Section~\ref{sec:modelCal}. The real-time electricity price used in off-line learning and in real-time control is a scaled version of PJM's locational margin price during Sept. 6-12, 2021~\cite{PJM:ElecPriceUrl}. Other relevant simulation parameters are located in Table~\ref{tab:simParams}. The policy evaluation problem~\eqref{eq:polEval} is solved using CVX~\cite{cvx}. The optimization problems to update the policy and the \plant\ dynamics are solved using CasADi/IPOPT~\cite{Andersson2019,wacbie:2006}.   

\emph{We emphasize that the closed loop results presented here are ``out-of-sample'' results, meaning the external disturbance $w_k$ used in the closed loop simulations are different from those used in training the RL controller.}
        
\subsection{Numerical Results and Discussion}
\begin{figure}[h]
		\centering
                \includegraphics[width=1\linewidth]{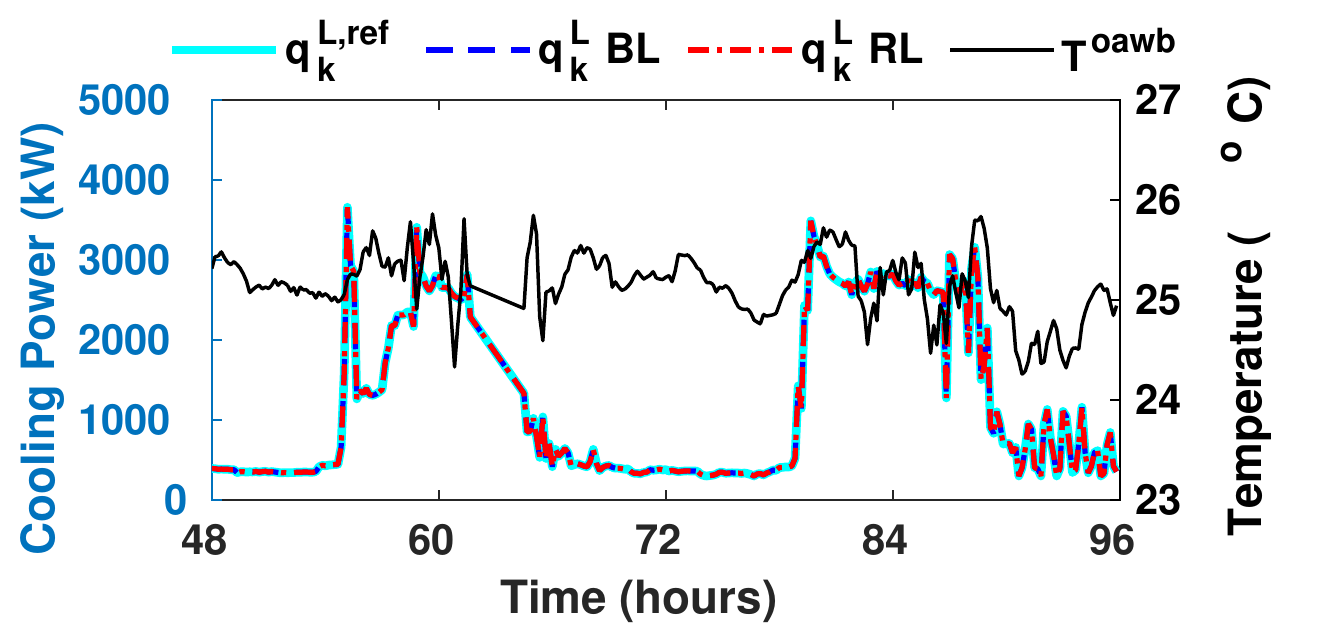}
		\caption{Load tracking performances for the RL and baseline controllers.}
		\label{fig:q_L}
	\end{figure}

Both the RL controller and the baseline controller meet the cooling load requirement, see Figure~\ref{fig:q_L}. But the RL controller outperforms the baseline controller. The total electricity cost of the RL controller - \$2,051/week - compared to that of the baseline controller - \$2,214/week, reaches a saving of 8\%.  For comparison, savings by MPC controllers with mixed-integer formulation reported in the literature are 9.7\% in~\cite{RisbeckMixedintegerEnB:2017} and 10.8\% in~\cite{DengMPCASE:2014}.

Simulations are done for a week, but the plots below show only two days to avoid clutter. 
\begin{figure}[h]
	\centering
	\includegraphics[width=1\linewidth]{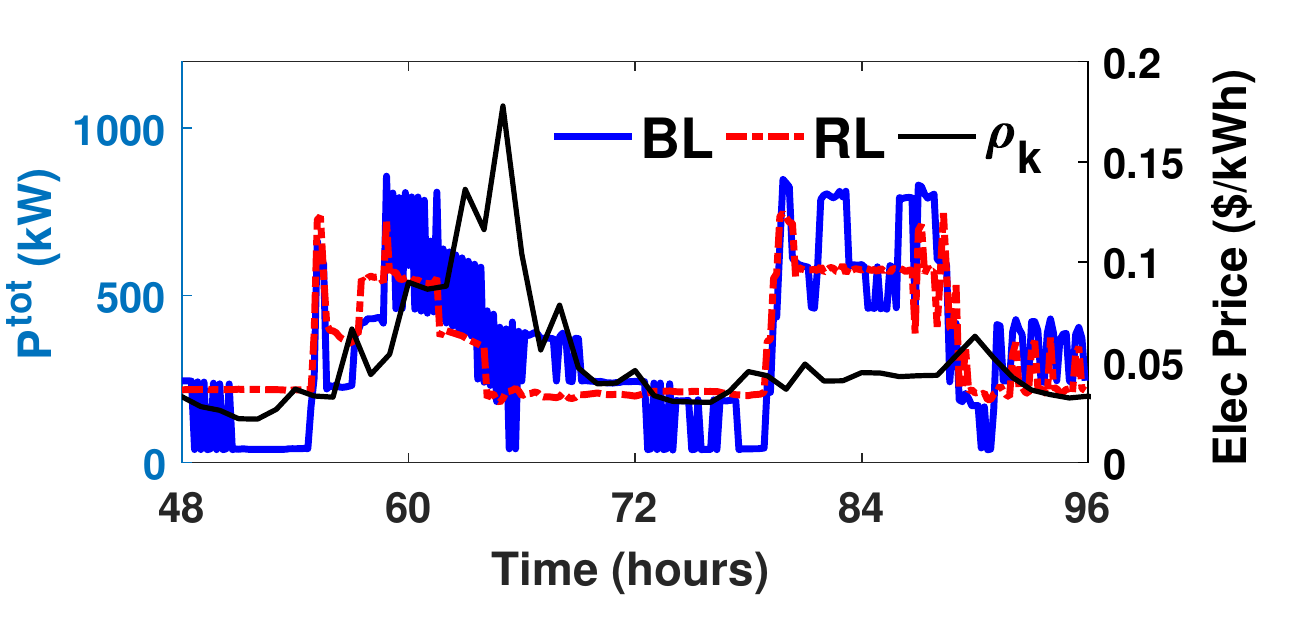}
	\caption{Power consumption and real-time electricity price.}
	\label{fig:P_tot}
\end{figure}

\begin{figure}[h]
		\centering
        \includegraphics[width=1\linewidth]{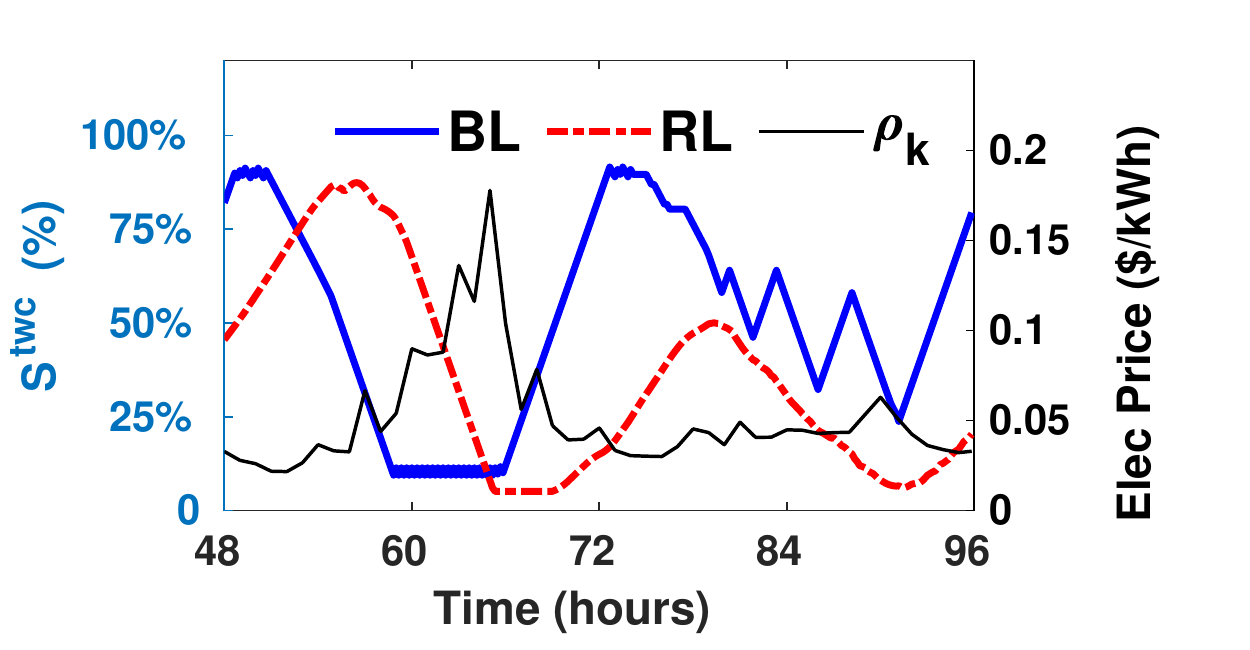}
		\caption{TES cold water volume along with electricity price. }
		\label{fig:mDot_TW}
	\end{figure}

      The cost savings by the RL controller comes from its ability to use the TES to shift the peak electric demand to periods of low price better than the baseline controller; see Figure~\ref{fig:P_tot}.         The cause for this difference is that the RL controller learns the variation in the electricity price well, or at least better than the BL controller.  This can be seen in Figure~\ref{fig:mDot_TW}. The RL controller always discharges the TES ($S^{\twc}$ drops) during the peak electricity price while the baseline controller sometimes cannot do so because the volume of cold water is already at its minimum bound. The BL controller discharges the TES as soon as the electricity price rises, which may result in insufficient cold water stored in the TES when the electricity price reaches its maximum. While both the controllers are forced to use the same price information (current and a moving average), the rule-based logic in the baseline controller cannot use that information as effectively as RL.        

\begin{figure}[h]
		\centering
                \includegraphics[width=1\linewidth]{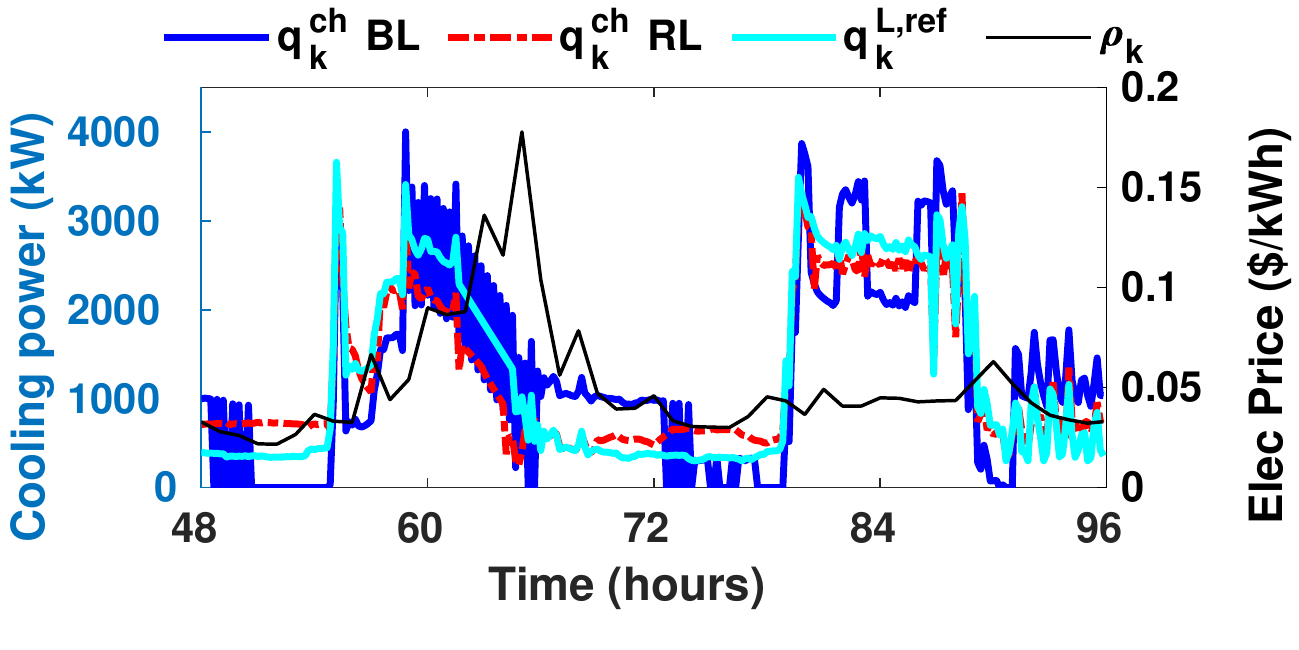}
		\caption{Cooling provided by the chillers, cooling load,  and real-time electricity price.}
		\label{fig:q_CH}
	\end{figure}

        An alternate view of this behavior can be obtained by looking at the times when the chillers are turned on and off and the resultant supplied cooling, since using chillers cost much more electricity than using the TES, which only needs a few pumps. We can see from Figure~\ref{fig:q_CH} that both BL and RL controllers shift their peak electricity demand to the times when electricity is cheap. But the rule-based logic of the BL controller is not able to line up electric demand with low price as well as the RL controller does.

\begin{figure}[h]
		\centering
		\includegraphics[width=1\linewidth]{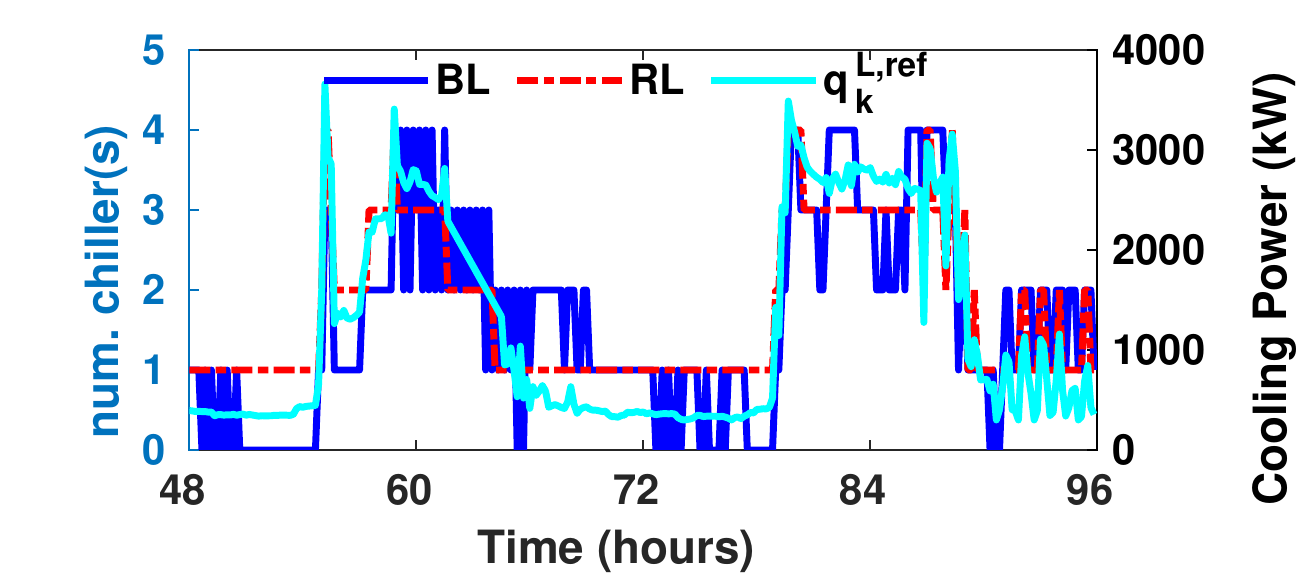}  
		\caption{Number of active chillers with respect to real-time electricity price for rule-based controller and RL controller}
		\label{fig:nChiller}
	\end{figure}

	Another benefit of the RL controller is that it cycles the chillers less than the BL controller even the cost of switching between on-off status of chillers is not incorporated in the cost function; see Figure~\ref{fig:nChiller}. Fast cycling decreases the life expectancy of a chiller greatly.
	\subsection{Lessons learned}  
\begin{figure}
		\centering
		\includegraphics[width=0.72\columnwidth, height=0.62\columnwidth]{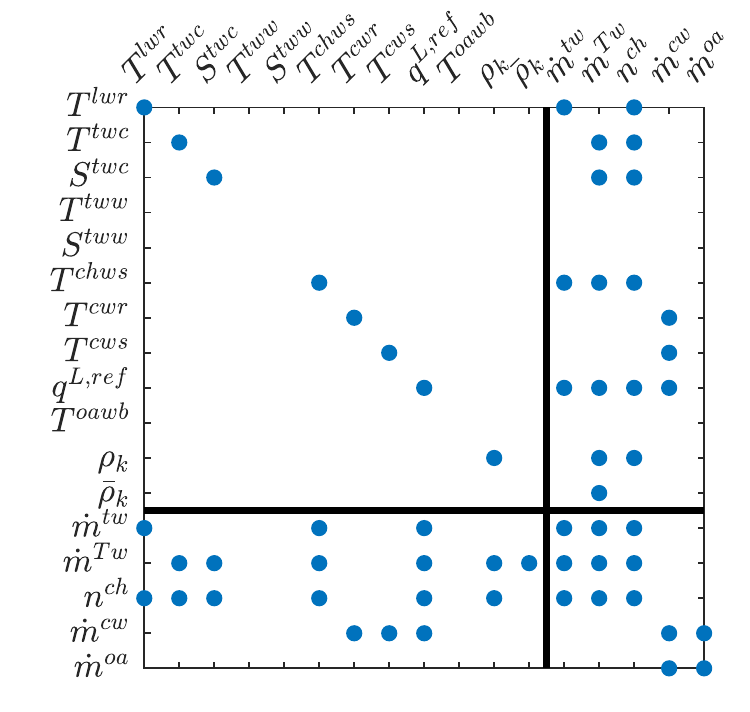}
		\caption{Sparsity pattern of the matrix $P_\theta$ appearing in~\eqref{eq:QfuncMatForm}.}
		\label{fig:sparPatternPtheta}
	\end{figure}

	Training of the RL controller is an iterative task that required trying many various configurations of the parameters appearing in Table~\ref{tab:simParams}. In particular,
	\begin{enumerate}
		\item If the value of $\kappa$ is too small, the controller will not learn to track the load $\qLref_k$. On the hand, if $\kappa$ is too large the controller will not save energy cost.
		\item The choice of basis defines the approximate Q-function, and consequently the policy~\eqref{eq:uk-argMinQ}. Redundant basis functions can lead to overfitting, which causes poor  out-of sample performance of the policy. We avoid this effect by selecting a reduced quadratic basis, which are the 35 unique non-zero entries in Figure~\ref{fig:sparPatternPtheta}. These choices were made based on physical intuition, e.g., the term $\mDotOA\times\mDotTW$ is not included as $\mDotOA$ and $\mDotTW$ have minimal dependency. 
		\item The condition number of~\eqref{eq:polEval} significantly affects the performance of Algorithm~\ref{alg:dataDrivePolIter}. The relative magnitudes of state and input values fundamentally determines the condition number. With appropriate scaling of the states/inputs, we reduce the condition number from $1\times 10^{20}$ to $1\times 10^3$. 
	\end{enumerate}      

	\section{Conclusion}\label{sec:conclusion}
     The proposed RL controller is able to reduce energy cost by 8\% over the baseline for a district cooling plant. The savings are comparable to that by MPC reported in other works while only using a fraction of MPC's real-time computational cost. 
     
     In future work, we would like to further test the robustness of our controller under different disturbance trajectories. We would also like to explore non-linear bases, such as neural networks, and examine convergence of the learning algorithm both numerically and through analysis.
	\bibliographystyle{IEEEtran}

\ifshowArxiv

    \appendix

\subsection{Simulation model of a \plant}\label{sec:sim-model}
Consider a multi-chiller system with a primary-secondary pumping, a water thermal energy storage tank, a set of cooling water pumps, and a cooling tower, as is shown in Figure~\ref{fig:ChillerPlant}. We now describe the plant states and their dynamics, along with the outputs. The \plant\ is described in two parts: (i) the chilled water loop and (ii) the cooling water loop. 
\subsubsection{Chilled water loop}\label{sec:sysModel_CHW}
The inputs to the chilled water loop are: \mDotLW, \mDotTW, and \nCh. Starting from the load, the load water supply is a mixture of chilled waters from both chillers and the TES.  Temporarily using continuous time, and time $t_k$ corresponding to the index $k$, the rate of enthalpy change of load water supply follows: (dropping $C_\pw$ from both sides) 
\begin{align} \label{eq:firstLawLWS}
	\frac{d}{dt}\Big(s^{\lw}(t)T^{\lws}(t)\Big)&\Bigg\vert_{t = t_k} \\ &= \begin{cases} 
		\dot{m}^{\sw}_kT^{\sw}_k - \dot{m}^{\tw}_kT^{\sw}_k, & \dot{m}^\tw_k > 0 \\
		\dot{m}^{\sw}_kT^{\sw}_k - \dot{m}^{\tw}_kT^{\twc}_k, & \dot{m}^\tw_k < 0 
	\end{cases}. \nonumber
\end{align}
Using the chain rule to expand the above differential, and supposing $T^{\lws}$ reaches steady state instantly (which yields $\dot{T}^{\lws} = 0$), we have from~\eqref{eq:firstLawLWS} that:
\begin{align}\label{eq:T_lws}
	T^{\lw,s}_{k} = T^{\sw}_k + \frac{\text{min}(\dot{m}^\tw_k, 0)}{\dot{m}^\lw_k}\Big(T^{\sw}_k - T^{\twc}_k\Big).
\end{align}
The load water return temperature $T^{\lw,r}$, by modeling through the heat balance, evolves as:
\begin{align}
	{T}^{\lwr}_{k+1} &= \frac{1}{C_{\pw}\dot{m}^{\lw}_k}(q_k^L + C_{\pw}\dot{m}_k^{\lw}T_k^{\lws}),\label{eq:T_lwr}\\
	0 &\leq q^L_k \leq q^{L, \text{ref}}_k, 5 \leq{T}^{\lwr}_k \leq 15 \label{eq:CCHeatBalance}.
\end{align}
The unit delay between the left hand side and the right hand side is to model the non-instantaneous nature of heat exchange between water and air streams, and the transport delay in the chilled water distribution pipes.  

The water thermal energy storage tank is modeled by two sub-tanks: warm water tank and cold water tank. This approach of modeling water TES is natural since a thermocline separates the warm water and cold water in the TES, which is the bedrock for TES to function properly~\cite{ASHRAE_TESguide:2019}. Associated with each tank are two variables: the temperature of the water, $T^\tww_k$ and $T^\twc_k$, and the mass of water, $s^\tww_k$ and $s^\twc_k$. The input for both tanks is the mass flow rate $\dot{m}_k^\tw$, which represents the water flow rate of two tanks. When $\dot{m}_k^\tw > 0$ this means that at time $k$, $\dot{m}_k^\tw t_s$ (kg) of water is charged into the cold water tank and $\dot{m}_k^\tw t_s$ (kg) has been discharged from warm water tank.  The dynamics for the storage variables are:
\begin{align} \label{eq:S_TW}
	s^\twc_{k+1} = s^\twc_k + t_s\dot{m}^\tw_k \quad \text{and}  \quad 
	s^\tww_{k+1} = s^\tww_k - t_s\dot{m}^\tw_k. 
\end{align}
Both quantities are constrained as $s^\tww_k, s^\twc_k \in [s^\text{min},s^\text{max}]$ with $s^\text{min} > 0$. The TES is considered well insulated, so that the temperature of the tank is only effected by the temperature of water coming into and out of the tank. For the warm water tank, heat balance reads:
\begin{align} \nonumber
	\big(T^\tww_{k+1}s^\tww_{k+1}-T^\tww_{k}s^\tww_{k}\big) = \begin{cases}
		-t_s \dot{m}_k^\tw T^{\lw,r}_k & \dot{m}_k^\tw < 0 \\
		-t_s \dot{m}_k^\tw T^{\tww}_k & \dot{m}_k^\tw > 0
	\end{cases}
\end{align}
which can be combined into one equation as:
\begin{align}\label{eq:T_tww}
	T^\tww_{k+1} = T^\tww_{k} + t_s \frac{\text{min}(\dot{m}^\tw_k,0)}{s^\tww_k - t_s\dot{m}^\tw_k}\Big(T^\tww_k - T^{\lw,r}_k\Big).
\end{align}
The cold water tank derivation is symmetric, and the final result is
\begin{align}
	T^\twc_{k+1} = T^\twc_{k} + t_s \frac{\text{max}(\dot{m}^\tw_k,0)}{s^\twc_k + t_s\dot{m}^\tw_k}\Big(T^{\sw}_k - T^\twc_k\Big).
\end{align}

The supply water flow rates are obtained once inputs $\dot{m}^{\lw}_k$ and $\dot{m}^{\tw}_k$ are chosen. The supply water temperature $T^{\sw}_k$, return water temperature $T^{\rw}_k$, and flow rate $\mDotSW$ are then:
\begin{align}
	T^{\sw}_k &= T^{\chws}_k \label{eq:T_sw}, \quad {\dot{m}}^{\sw}_k = \dot{m}^{\lw}_k + \dot{m}^{\tw}_k, \ \text{and} \\
	T^{\rw}_{k} &= T^{\lw,r}_k + \frac{\text{max}(\dot{m}^\tw_k, 0)}{\dot{m}^\sw_k}\Big(T^\tww_k - T^{\lw,r}_k\Big).	
\end{align}

In a primary-secondary pumping system, the water goes through each active (``on'') chiller is a constant $\dot{m}^{\indv}$. The total chilled water produced, $\dot{m}^{\chw}_k$, may be more than what is required by loads and TES; a one-way bypass valve sends the redundant chilled water, $\dot{m}^{\bp}_k$, from chiller outlet to chiller inlet. Chilled water return $T^{\chwr}_k$ is a mixture of return water and bypass water. The above relationships are summarized as:
\begin{align}
\dot{m}^{\chw}_k &= n^{\ch}_k \dot{m}^{\indv}, \ \dot{m}^{\bp}_k = \dot{m}^{\chw}_k - \dot{m}^{\sw}_k, \ \dot{m}^{\bp}_k \geq 0 \label{eq:mDot_BP}, \\
T^{\chwr}_k &= T^{rw}_k + \frac{\dot{m}^{\bp}_k}{\dot{m}^{\chw}_k}( T^{\chws}_k-T^{\rw}_k ). \label{eq:T_chwr}
\end{align}
Assume all chillers are identical and each chiller has a nominal cooling capacity of $q^{\indv}$, then the chilled water supply temperature $T^{\chws}_{k+1}$ coming out of the chiller evaporator is:
\begin{align}	
	{T}^{\chws}_{k+1} &= T^{\chwr}_{k} -  \frac{\dot{q}^{\ch}_k}{C_{\pw}\dot{m}^{\chw}_k},   \label{eq:CHHeatBalance} \\
	\dot{q}^{\ch}_k &\leq n^{\ch}_k \qChCapIndiv,  \quad	\TChWS_k  \geq \TChWSMin.  \label{Cons:T_chws}
 \end{align}
\subsubsection{Cooling water loop} \label{sec:sysModel_CW}
The inputs to the cooling water loop are: \mDotCW and \mDotOA. Simply speaking, without elaborating on the refrigerant loop in chillers, part (or all) of the heat absorbed by the chilled water at the buildings is removed at the chiller and then injected into the cooling tower (supply) water, which reduces the chilled water's temperature from $T_k^{\chwr}$ to $T_k^{\chws}$ and increases cooling water's temperature from  $T_k^{\cws}$ to $T_k^{\cwr}$. The above heat exchange occurs at the chiller condenser. The rate of this heat exchange is denoted by $\dot{q}_k^{\cond}$, with the superscript denoting condenser, and is modeled via heat balance: 
\begin{align}
	\dot{q}_{k}^{\cond} = \dot{q}^{\ch}_k + \eta_1P_k^{\ch},
\end{align}
where $\eta_1P_k^{\ch}$ is the waste heat from the chiller compressor motors, and $P_k^{\ch}$ is described in~\eqref{eq:P_CH}. Following the heat balance, $\dot{q}_{k}^{\cond}$ results in an increase of the cooling water:
\begin{align}
T_{k+1}^{\cwr} & = \frac{\dot{q}_k^{\cond}}{C_{\pw}\dot{m}_k^{\cw}}+T_k^{\cws}, \quad T_{k}^{\cwr} \leq \TCWRMax.
\end{align}
The cooling capacity of a cooling tower, denoted $\dot{q}_k^{\ct,\rej}$, is modeled as~\cite{JinSimplifiedECM:2007}:
\begin{align}
\label{eq:q_CT_rej} 
\dot{q}_k^{\ct,\rej} = \frac{c_1 (\dot{m}_k^{\cw})^{c_3}}{1 +  c_2 \left( \frac{\dot{m}_k^{\cw}}{\dot{m}_k^{\oa}}\right)^{c_3}}(T_k^{\cwr}-T_k^{\oawb}),
\end{align}
where $c_1,c_2,c_3$ are empirical constants. The heat rejected through evaporation at the cooling tower is denoted ${\qDotEvap}_k$,  and is modeled from the heat exchange at the condenser by a unit delay:
\begin{align}
	\TCWS_{k+1} & = \TCWR_k - \frac{\qDotEvap_k}{C_{\pw}\mDotCW_k }, \quad 	\TCWS_{k+1}  \geq \TOAWB_k + 1, \label{eq:T_CWS}\\
	0 \leq &\qDotEvap_{k} \leq \dot{q}_{k}^{\ct,\rej}. \label{eq:CTHeatBalance}
\end{align}

In order to simulate the dynamics of the \plant, the capacity of all heat exchange devices, specified through the constraints in~\eqref{eq:CCHeatBalance},~\eqref{Cons:T_chws}, and~ \eqref{eq:CTHeatBalance},  need to be satisfied. 
An efficient way to ensure this is through a constrained optimization with a ``projection objective function.'' The decision variable $Z_k$ for the optimization problem consists of the next system states and the current outputs:
\begin{align}
	Z_k \triangleq \big[T^{chw,s}_k, T^{cw,s}_k, \ \qDotLoad_k\big]^T.
\end{align}
This variable $Z_k$ is solved for by projecting it onto the dynamics and heat exchange capacities of the system. The system then evolves as:
\begin{align} \label{eq:plantDyn}
	Z^*_k = \arg&\min_{Z_k\in\Omega(x^p_k,u_k)} \| Z_k - \bar{Z}_k\|_A,
\end{align}
where $\bar{Z}_k$ is a vector of user-specified setpoints to reflect system nominal working status. Specifically, $\bar{\dot{q}}^L_k$ = $\dot{q}^{L,\text{ref}}_k$. Diagonal matrix $A$ is to set a trade-off between maintaining $x_{k+1}^p$ at nominal working status and ensuring cooling capacities are respected; set $\Omega(x_k^p,u_k)$ is defined by the dynamics and constraints of \plant\ system:
\begin{align} \nonumber
\Omega(x_k^p,u_k) \triangleq\big\{\mathbf{Z} : \mathbf{Z} \text{ satisfies} \eqref{eq:P_CH} - \eqref{eq:P_tot} \text{ and} ~\eqref{eq:T_lws}-\eqref{eq:CTHeatBalance}\big\}.
\end{align}

\end{document}